# Extracting, Visualizing, and Learning from Dynamic Data: Perfusion in Surgical Video for Tissue Characterization

Jonathan P. Epperlein
*IBM Research Europe*
Dublin, Ireland
https://orcid.org/0000-0003-2912-0543

Niall P. Hardy
*UCD Centre for Precision Surgery,*
*School of Medicine*
*University College Dublin*
Dublin, Ireland
https://orcid.org/0000-0002-7036-3910

Pol Mac Aonghusa
*IBM Research Europe*
Dublin, Ireland
https://orcid.org/0000-0002-7640-9668

Ronan A. Cahill
*UCD Centre for Precision Surgery,*
*School of Medicine*
*University College Dublin*
Dublin, Ireland
https://orcid.org/0000-0002-1270-4000

*Abstract*—Intraoperative assessment of tissue can be guided through fluorescence imaging which involves systemic dosing with a fluorophore and subsequent examination of the tissue region of interest with a near-infrared camera. This typically involves administering indocyanine green (ICG) hours or even days before surgery and intraoperative visualization at the time predicted for steady-state signal-to-background status. Here, we describe our efforts to capture and utilize the information contained in the first few minutes after ICG administration from the perspective of both signal processing and surgical practice. We prove a method for characterization of cancerous versus benign rectal lesions now undergoing further development and validation via multicenter clinical phase studies.

## I. Introduction

On the spot, in-surgery assessment of tissue (such as ascertaining the presence or extent of cancer or judging sufficiency of tissue perfusion to ensure subsequent uncomplicated healing) remains one of the biggest challenges to the modern surgeon who, despite significant advances in pre-operative imaging and surgical instrumentation, remains effectively unaided in these determinations. The deployment of fluorescence guidance, utilizing the fluorophore indocyanine green (ICG) with a near infrared (NIR) imaging system, represents an exciting recent development through which operators can glean further information by performing real time non-invasive and radiation-free angiograms and lymphangiograms that have also been shown to demarcate malignant deposits in the tissue by secondary non-selective uptake of the dye into the neoplastic tissue by passive and active cancer cell accumulation [2][3][4][5].

The clinical usefulness of fluorescence guidance for either tissue healing or cancer determination has not yet been fully established however although there are increasing numbers of clinical and academic groups working in this area. While NIR imaging for perfusion assessment of the large intestine at the time of resection and re-joining (anastomosis) has generally (but not uniformly) shown positive benefit in reducing non-healing (leak) rates, it remains at present a subjective determination in that the operator still has to judge the on-screen appearances [6][7]. Interobserver variability in operator interpretation of the subtle and complex fluorescence changes over time has been previously reported especially among inexperienced users [8][9][10]. Fluorescence guidance for cancer detection is even less established with only very limited indications and then only in certain surgical centers [11][12]. The current state of the art for cancer delineation involves the administration of ICG hours/days in advance of surgery and with NIR imaging then used intraoperatively to detect the dye's presence within malignant tissues (aiming for the window in which dye retention in the region of interest is maximal while persistence in other tissue is minimal). This approach has suffered poor specificity and presents organizational challenges given the need for precursory dosing [13][14].

With confidence in the fundamental, biophysical discriminatory principle of ICG flow and impedance through tissues, we sought to apply modern computer vision and artificial intelligence methods to enable the dynamic assessment of tissue fluorescence immediately following intravenous ICG administration to supplement human interpretation of perfusion angiograms intraoperatively [15][16]. For cancer characterization and margination, such capability could too replace static, single-point-in-time imaging in favor of a rapidly deployable intraoperative adjunctive imaging tool especially if on-screen overlay of discriminatory, additional information could be achieved. This paper details our successes (both technical and clinical) in doing so and also the obstacles encountered and remaining in our efforts to shift on the current state of the technological and operative art of complex, intraabdominal minimally invasive cancer surgery.

*Current State of the Art.* The surgeon intending fluorescence guided surgery uses a clinically approved digital camera and NIR imaging system to examine tissues intraoperatively. The fluorophore is administered intravenously during such observation and the surgeon makes a judgement of the adequacy of the dye flow in real time over a period of seconds to a few minutes. Alternatively, in the case of cancer

This work was partially supported by Disruptive Technologies Innovation Fund, Ireland, project code DTIF2018 240 CA

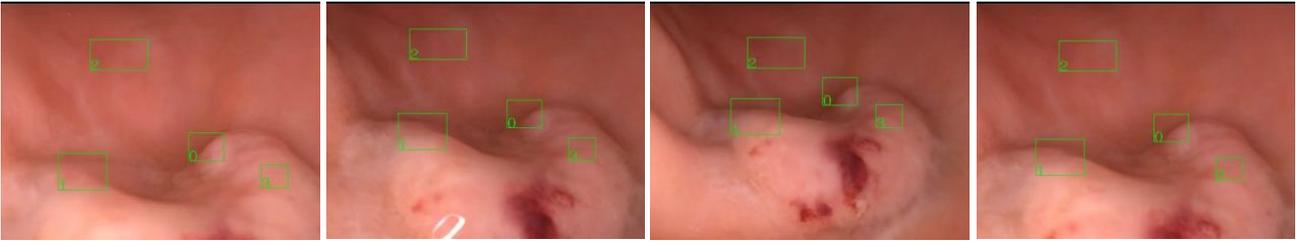

*Figure 1: The region tracking algorithm keeps the ROIs selected in the first frame attached to the same region of tissue, despite considerable movement.*

localization, the dye may have been given sometime preoperatively and the surgeon now uses the NIR camera to examine for fluorescence retention to supplement standard tissue examination (in white light). To apply computer vision and machine learning augmentation to these processes to move beyond the current state, several challenges are encountered needing a combination of clinical protocolization and technological optimization to address.

## II. Signal Processing of Endoscopic Video

Using endoscopic video, collected in a real-world, intraoperative setting, as a source for dynamic, i.e. time-resolved, data presents two basic technical challenges. The first is intrinsic to the process of image acquisition: endoscopes are typically handheld, and patients are either conscious or anesthetized, so that motion, deformation, and occlusion occur and have to be compensated for prior to data collection. A collected time series should reflect the fluorescence over time in a specific region of tissue, and not at a specific pixel location in the video. While of course the camera can be held relatively steady at a standard distance from the tissue, motion still occurs at both the camera and tissue level. To this end, we developed, evaluated, and released a number of computer vision algorithms, which we briefly describe in II.A.

The second challenge is with the functional nature of that data: for each area of observation, which can be a region of interest, or each individual pixel, the data is a collection of time series. Due to inter-patient variability, inconsistent protocols of ICG administration, and loss from view, the time series can be of quite uneven lengths. Summarizing this data into either a feature vector for subsequent machine learning or as an additional source of information in the surgeon's decision is a non-straightforward problem, and we go into this in some detail in II.B

### A. Computer Vision

Motion compensation can be accomplished by selecting certain regions of interest (ROIs) and algorithmically tracking their location throughout the video. This is conceptually related to classical object tracking, which is a mature topic in computer vision, but it turns out that due to the often smooth textures and the lack of "objectness" of ROIs (see e.g. ROI 3 in Figure 1), most off-the-shelf object tracking algorithms perform poorly [24]. We developed a simple region tracking algorithm based on optical flow with outstanding performance on endoscopic video at framerates of 30 FPS (on a MacBook Pro at 480x360 resolution). We have released our Python implementation as open source [24][25] and used it ourselves in our previous publications e.g. [23][16][17][15][1].

The bounding box for each ROI is translated from the visible light image to the NIR image, and the brightness within it is taken as the measurement. This process yields one time series per selected ROI. This would allow for classification of the tissue within each ROI, and like a biopsy, it requires the surgeon to make a decision regarding which region to probe. For regions that are not selected, no information is gained (some authors use the term "optical biopsy" to emphasize the analogy). Typically, spatial relationships between regions are not considered in this framework, each ROI is taken as a monolith, and constraints such as "close-by ROIs should have similar labels" are not enforceable. These drawbacks – the manual selection of ROIs, the loss of information outside ROIs, and the discarding of spatial relationships – led us to consider another approach to remove motion from the video: estimating the entire deformation field, i.e. a function $U(x,y,t)$ so that the pixel located at *(x,y)* in frame 0 is moved to location $(x+U_1(x,y,t), y+U_2(x,y,t))$ in frame *t*. Applying this deformation in reverse, we obtain a stabilized video. More specifically, a stabilized video is generated by taking the pixel at location $(x+U_1(x,y,t), y+U_2(x,y,t))$ in frame *t* and placing it at location *(x,y)* in the *t*-th frame of the stabilized video. Figure 2 demonstrates this on a real example. There, the visible light channel is used to estimate *U*, but the same deformation is applied to the NIR channel, to stabilize both images. If $(x+U_1(x,y,t), y+U_2(x,y,t))$ is a non-integer value, a step of interpolation from the neighboring pixels is necessary, and if it lies outside the frame, no data is available for *(x,y)*; the green areas in Figure 2 correspond to this "no data" case.

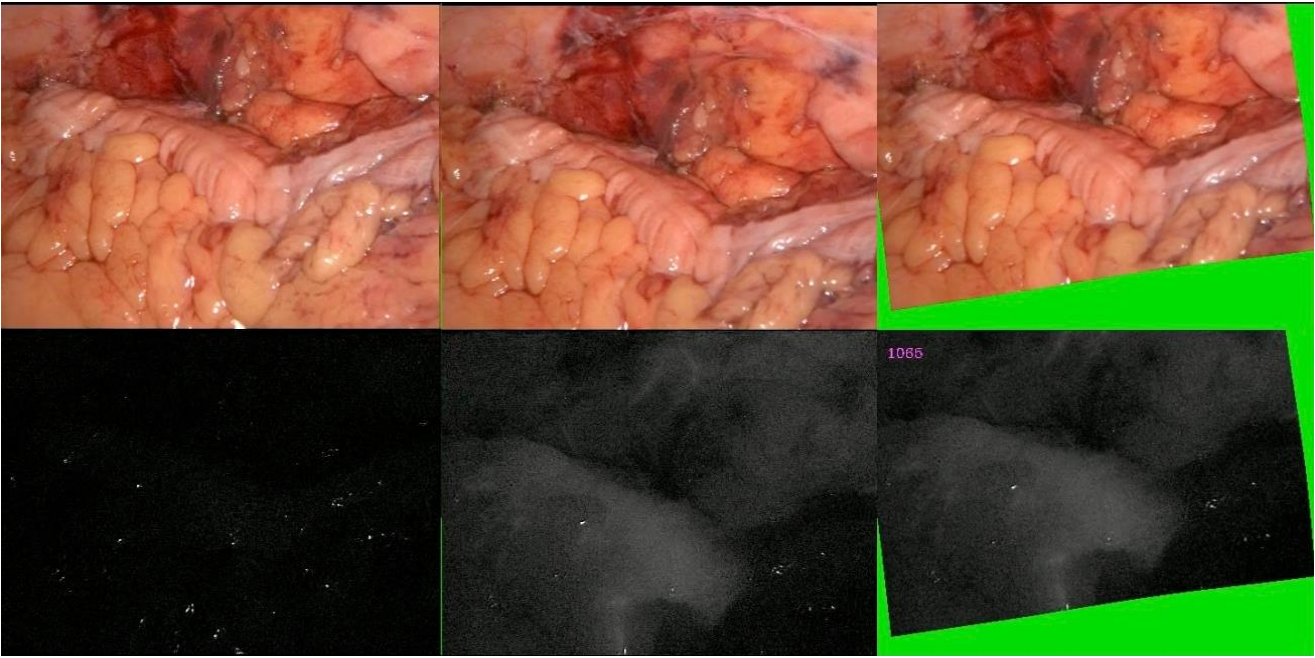

*Figure 2: From left to right: (1) The first frame of a video, where the top panel is visible light, and the bottom panel is the NIR channel. (2) A subsequent frame, where some movement and slight deformation has occurred. (3) The same frame, but the motion and deformation has been compensated, effectively moving the colon back into its original location within the video; green indicates areas outside the frame, hence no data is available there.*

Computationally, this is considerably harder than tracking a few ROIs. Our approach relies on the detection of key points in each pair of frames. Those key points are then matched, a process by which a point in frame 0 is identified with its new location in frame *t*, which yields a sparse set of values for the deformation field. These values are then interpolated with thin-plate splines. We've implemented our approach in a modular Python package, allowing for different type of key point, matching, and interpolation algorithms to be swapped in and out seamlessly. A fully functional, but very early-stages open-source release is planned to be released shortly.

The data so extracted is a full *field* of measurements, as the NIR panel of the stabilized video now contains a measurement *I(x,y,t)* for each pixel *(x,y)* and each frame *t* (some of which might be "no data", as explained above). The spatial resolution of the data means that it can be used for more advanced tasks, such as image segmentation.

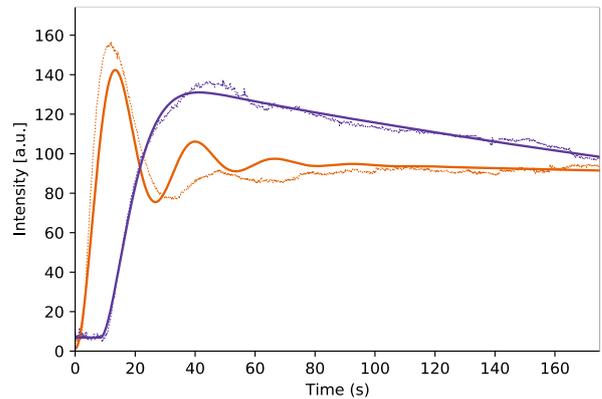

*Figure 3: Raw data (dotted lines with bands) and their fits according to (1) (solid lines).*

### B. Data Processing

The data obtained using the techniques outlined in the previous section take the shape of either a few time series $I_{p,r}(t)$, indexed by patient *p* and ROI number *r*, or fields $I_p(x,y,t)$ indexed by patient only, with *(x,y)* denoting the pixel location in the first frame. In the rest of the section, we explore different ways of using the data in a clinical context.

#### 1) 1-D Data

Learning from time series is more complex than learning from fixed-length vectors: techniques such as LSTMs, RNNs, or even HMMs typically require large amounts of data and yield classifiers whose results are hard to explain; more recent approaches such as MiniRocket [26] require smaller amounts of data, but still aren't "explainable". Our goal was to find a pipeline which performs well (obviously) yet doesn't require large amounts of data and reaches its classifications in a transparent, explainable way, based on meaningful features.

In [3][15], inspired by works like [19], we explored the use of parametric models. Namely, a simplified 2-compartment model (capillaries/veins and tissue) with exponentially decaying ICG input (from a 3rd, arterial compartment) is modeled as a 2nd order differential equation

$$\tau^2 \ddot{y}(t) + 2D\tau \dot{y}(t) + y(t) = Ke^{-\frac{t}{\tau_i}}, \quad (1)$$

where *y* denotes the observed brightness (which is assumed to be proportional to a weighted combination of the ICG concentrations in all three compartments), the dot denotes derivation with respect to time, and the parameters $(D, \tau, \tau_i, K)$, along with a constant for background fluorescence and a time delay to account for the unknown moment of ICG administration, are estimated. This set of 6 parameters can then be used as the feature representation of the time series. We used this feature set, along with a few derived features, to train a collection of explainable classification algorithms, including Nearest Neighbors, Naive Bayes, and Decision Trees, and achieved encouraging Accuracy/Sensitivity/Specificity of over 90% for the two-class discrimination (cancer vs benign/healthy) on a limited data set (20 patients).

The features based on the 2nd-order parametrization had two major drawbacks. Since the parameter estimates are "global" in the sense that each parameter depends on the entirety of the time series, they turned out to be sensitive with respect to the amount of data collected: the same time series truncated earlier (e.g. the curves in Figure 3 truncated at 100s) can result in very different parameter estimates. Clearly this is an issue. And on the other hand, while it is clear for most from an engineering background what the "damping $D$" or "time constant $\tau$" signify, it is much less intuitive to a medical practitioner, hence the features are physically inspired, or "explainable" to some, yet they are quite inaccessible to the professionals ultimately basing their decisions on them.

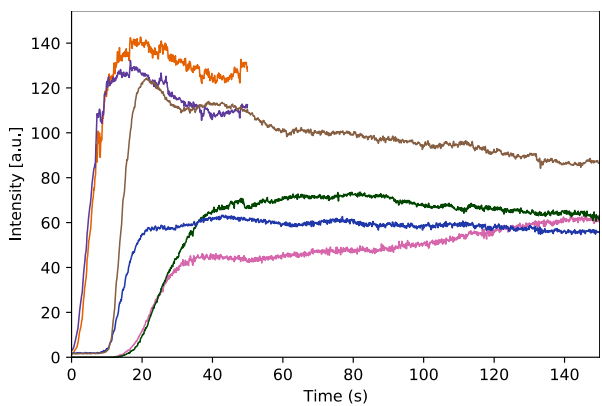

*Figure 4: The collected curves of fluorescence are of varying lengths and shapes, which is dictated by patient physiology, ICG administration protocols and video quality.*

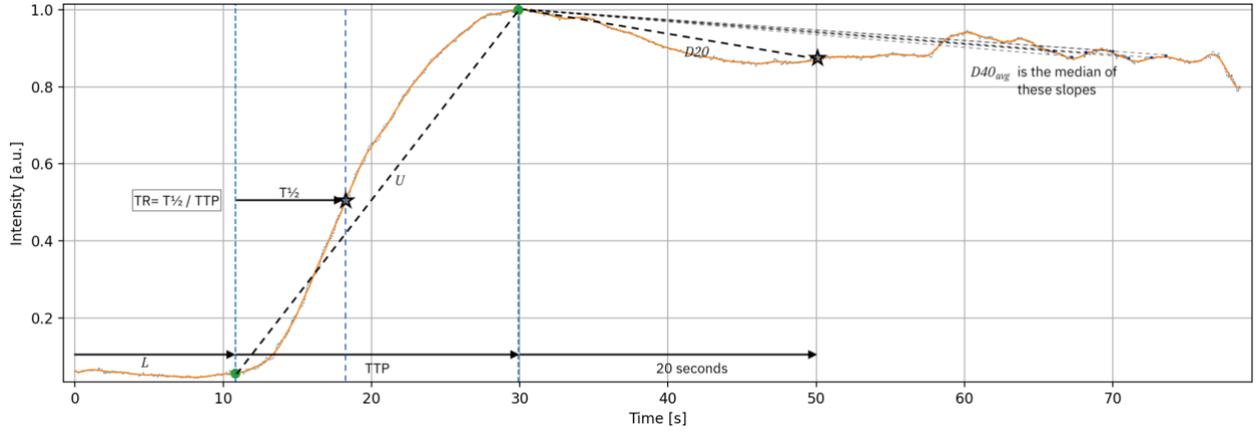

*Figure 5: An example of raw data (dotted), its smoothed version (solid) and an illustration of the "simple features". This figure first appeared in [17].*

To address both disadvantages, in [17] we considered "simple features", i.e. characteristic values that are easy to visualize and even read off manually (of course we used signal progressing techniques instead), such as the time of the peak ("TTP") or several approximate slopes, see Figure 5 for illustrations. Every time series is then represented by its set of time to peak, up- and downslopes, and the "time ratio" TR. We trained the same collection of explainable classification algorithms on our data set, but this time sequentially decreased the number of features until only 2 features remained, which allows us to show decision boundaries for "ultimate explainability", again obtaining Accuracy/ Sensitivity/ Specificity between 85% and 95%, with the same caveat of very limited data.

*2) 2-D Data*

While it would be conceptually easy to simply consider each pixel as a ROI and apply the methods outlined in the previous section, that would discard the additional spatial information we now have and increase the scale of the problem by adding many redundant data points (two adjacent pixels will most of the time have almost identical time series). While we have not yet fully explored methods utilizing such field-valued data, we have investigated visualizing one or more characteristic values, such as the ones outlined in Figure 5, in a heatmap, basically providing the surgeon with an additional source of information, an additional dimension to the data, or a "virtual modality".

An example of characteristic value that has proven useful is the "center of mass"

$$\mu = \delta t \sum_{k=0}^{N} k \cdot y[k] \Big/ \sum_{k=0}^{N} y[k]$$

i.e. the time point on which the intensity would "balance". Here, $N$ denotes the number of time steps, and $\delta t$ is the time between two consecutive frames, hence it's equal to the inverse of the frame rate. This helps to distinguish between regions of ICG accumulation (center of mass at later time points) and regions where brightness peaks and then decays. Any of the other characteristic values can also be visualized; Figure 6 contains examples.

*C. Current efforts and future work*

The current work streams can be summarized as "do more with the 2-D data". The most obvious direction is to apply more sophisticated machine learning techniques, such as image segmentation networks, to the field-valued time series, ultimately enabling us to draw application-specific heat maps. In a diagnostic procedure to detect cancer, the heat map would correspond to areas of suspicious growth; in anastomosis planning, the viability along the bowel could be estimated and displayed.

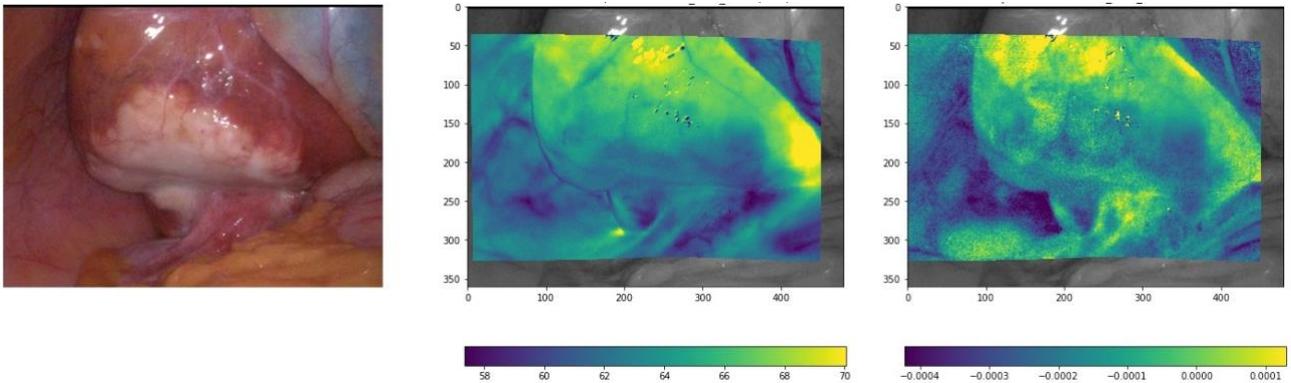

*Figure 6: Heatmaps overlaid on the initial frame (left). Center: the colormap corresponds to the first central moment μ of the time series, darker color indicates that more of the fluorescence happens early on, whereas the yellow hue indicates that ICG accumulates. Right: the colormap corresponds to the approximate decay slope between peak and 20 seconds past the peak. Positive slope indicates accumulation, negative slope indicates washout.*

Another, much more computationally expensive line of inquiry concerns the optical properties of the imaged tissue. The collected fields $y(x,y,t)$ are observations of a 3-D field $y(x,y,z,t)$ of photon density on the camera plane. With known camera and source locations, and a model of photon propagation (the "photon diffusion equation") containing the optical parameters of the tissue, a reconstruction of these properties can be attempted, an instance of diffuse optical tomography (DOT). The optical properties correlate with the biological properties, e.g. a high absorption coefficient at the ICG excitation wavelength (~790nm) indicates the presence of ICG, so this coefficient can be used as a proxy for the concentration of ICG, in 3-D! Many challenges corresponding to poor conditioning of the underlying numerics, and the sensitivity to the very noisy real-world data, still have to be overcome to make this a reality.

Lastly, the visualization of information as a static image can be improved upon by embedding it into an augmented reality view, basically virtually "tattooing" the additional information, such as slopes or classification confidences, onto the imaged tissue.

### III. CLINICAL APPLICATION

The computational capability described here has been applied to series of video recordings of tissue perfusion in real patients undergoing surgery for cancer and benign disease in three different university hospitals working under the same clinical protocol and indeed trial parameters. As the trial continues, increasing numbers of video recordings are available allowing incremental application of the developing methodology with feedback of the findings to inform iterative algorithmic evolution. In this way, the foundational discovery that fluorescence flow through the tissue dynamically informs regarding the tissue's nature (e.g. whether it is healthy or unhealthy or indeed whether any particular target within it is cancerous or not) by synchronous comparison of an area of abnormality to an adjacent area of normality within the same field of view has moved from initial proof of concept (achieved by plotting fluorescence time curves from selected sample frames as an image stack using ImageJ in a small series of patients with rectal neoplasia) to a corpus of work with evidence of generalized applicability (including a substantially greater number of patients with rectal neoplasia and a second group of those with colorectal liver metastases) as well as broadening to include patients from three different cancer hospitals in two different countries. The next step clinically is implementation in a broad validation study in several European countries including prospective interventional trialing culminating in a randomized control trial with biolegal research regarding perspectives in acceptability re patients, surgeons, healthcare providers and payers as well as professional surgical societies and guideline organizations.

To date we have seen high levels of accuracy in the computational classification of significant rectal polyps (polyps >2cm) as either cancerous or non-cancerous. This is a crucial distinction as the treatment and prognosis differs completely depending on whether the neoplasia inherent in a polyp is invasive (and therefore a cancer) or non-invasive vis a vis the epithelial layer. A benign lesion can be cured by complete local excision whereas a cancer needs radical excision including all of its draining lymphatic basin. For a rectal tumor this is the difference between an internal, organ sparing procedure and an invasive extirpative operation with major risks of non-healing (and so sepsis) and stoma formation in cases in which restoration of bowel continuity cannot be achieved. The clinical problem is that the true nature of such lesions is difficult to distinguish in advance of the actual excision. Methods to profile without excision include endoscopist opinion (based on phenotypic appearance as well as the observers own experience and knowledge), radiological assessment (CT, MRI and ultrasound), and pathological methods (biopsy), which are all imperfect for lesions of this size and in this area [21][22]. This means that some cancers are initially (under)treated as benign lesions while the converse is also true with some benign lesions being (over)treated as cancerous lesions. All investigations also take time and add expense to the healthcare journey of the patient and system. An ability so to define the nature and margins of the lesions instantly visually at endoscopic intervention would greatly acuminate healthcare and outcomes in this area.

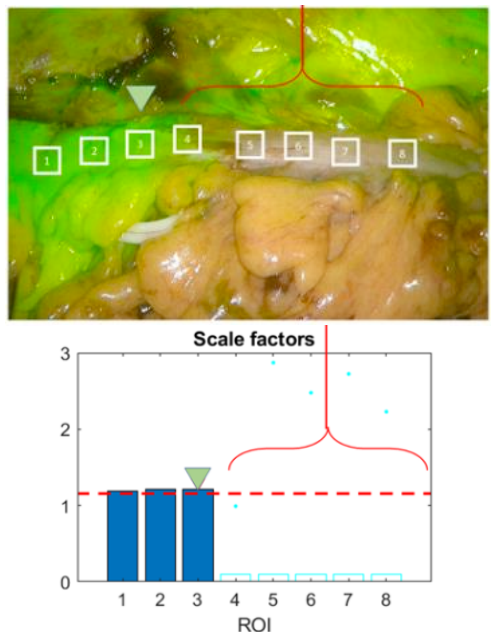

*Figure 7: Example of anastomosis planning as elaborated in [23]. Dynamic data as in e.g. Figure 4 is collected, and a temporal scale is estimated. Smaller scale factors suggest a curve is "stretched" more, which in turn suggests sluggish perfusion.*

Expanding on this, similar methodology has been deployed for characterization of liver metastases, secondary deposits from a colorectal cancer. For this use, the basic concept differs from rectal lesions in that the liver concentrates circulating ICG ahead of its excretion, different to other tissues which tend to clear the dye after an initial peak following administration. Therefore, the comparative clearance of the dye between the abnormal tissue and the normal tissue is reversed compared to that seen in the liver. In addition, we have observed similar discriminative capability in gynecological cancers as well as in lymphomatous deposits. Furthermore, we have now applied the computational methodology to assessment of perfusion of tissue after surgical excision at the time of the operation where the focus is on reconstruction for functional optimization. This applies to colorectal surgery (restoration of gastrointestinal continuity) as well as plastics and reconstructive surgery including breast cancer surgery. Here the crucial importance is in ensuring that the tissues being opposed are sufficiently healthy to heal together both in terms of baseline blood supply and any tensioning inherent in approximating tissues. This is the current main use of fluorescence imaging in surgery although the qualitative nature of such implementation persists as a limitation. Further work to quantify and analyze fluorescence signals in the setting of intestinal resection and anastomosis, including with comparison to control healthy tissue, capillary lactate levels and tissue oxygen saturation, is ongoing. In combination with multiple, sequential user-selected ROIs along the bowel in question, and ultimately automated 2D representation, such analysis may provide objective feedback regarding bowel perfusion capturing the dynamic nature of blood flow and, in time, even act as an intra-operative bowel transection recommender, see Figure 7.

For best use in all these applications, the added information needs to be displayed in a way that makes the surgeon's tasks simpler and easier rather than more complex. Cognitive burden is a considerable concern in surgery where at present the surgeon is entirely responsible both for all decisions involved in an operation (and any operation is a complex series of sequential decisions) and the functioning of the instrumentation. Heatmap overlay of appropriately thresholded data seems an appropriate method in surgery for guidance and initial experience with surgeons of different grades supports this. Of course, provision of such data in real-time during surgery (as opposed to post hoc on video recordings) requires high levels of computational capability but advances in edge and cloud processing indicates this to be technically achievable.

Lastly, real-world practice parameters need consideration. Healthcare is expensive and added technological capability in surgery contributes considerably to this. While robotic platforms are in many ways ideal platforms to showcase data augmentation displays, the capital investment currently required means that only 3% of surgeries are now performed in this way. Software adjuncts have the capability of being added on to existing imaging systems and therefore have the potential to right shift surgical outcomes wherever in the world minimally invasive surgeries are currently performed. Indeed, such decision-support systems may have biggest benefit in assisting surgical care where the expertise that comes from the privilege of specialization and centralization is impossible and instead practitioners are required to be generalists. For all these reasons, computational methods that supplement existing technologies that have already proven safe, affordable, and acceptable in terms of operative and hospital workflow are an exciting innovation for surgical care.